\documentclass[twoside]{article}
\usepackage{fleqn,espcrc2}

\usepackage{graphicx}
\usepackage{epsfig}

\usepackage[figuresright]{rotating}

\newcommand{\AmS}{{\protect\the\textfont2
  A\kern-.1667em\lower.5ex\hbox{M}\kern-.125emS}}

\newcommand{\chup}{$\chi U\phi$}

\title{Spinons and holons on the lattice}

\author{Ji\v{r}{\'{\i}} Jers\'ak \address{Institute 
of Theoretical Physics E, RWTH Aachen, D-52056 Aachen, Germany}%
}        
       
\begin{document}

\begin{abstract}
  We point out that the dynamical fermion mass generation in the 3D compact
  U(1) lattice gauge theory with charged fermion and scalar fields (\chup$_3$
  model) may be of relevance for the spinon-holon theory with local gauge
  symmetry in the condensed matter physics. However, many properties of the
  \chup$_3$ model are uncertain, so we make some conjectures to motivate their
  future investigation. Most probably, for strong gauge coupling the model is
  by universality equivalent to the familiar 3D four-fermion coupling models
  with $N_f=2$. Available numerical results indicate that at the intermediate
  and weak gauge coupling two universality classes with new interesting
  physics may arise. One of them is associated with a tricritical point which
  probably exists in the phase diagram of the \chup$_3$ model. The other one
  is determined by the dynamical fermion mass generation in the compact
  QED$_3$, which is insufficiently understood but of much interest by itself.
  \vspace{1pc}
\end{abstract}

\maketitle

\section{Controversial background}

One of the attempts to understand the high temperature superconductors in the
condensed matter physics is based on the strongly coupled 3D U(1) gauge
theory. The gauge symmetry arises essentially through the Hubbard-Stratonovich
transformation of the nearest-neighbour four-fermion coupling (see
ref.~\cite{Fr91} for an explanation), and the strongly coupled U(1) gauge
field is thus different from the electromagnetic field. Including the idea of
spin-charge separation, this U(1) gauge field couples to fermion and boson
fields (spinons and holons, respectively) and is naturally compact. The
kinetic term is absent in the original formulation, i.e. $\beta = 1/g^2 = 0$,
but may arise during a renormalization or if some degrees of freedom are
integrated out.

Whether this framework, originated by P.~W.~Anderson
\cite{An87}, is really suitable for achieving the original goal
is a highly controversial issue \cite{Localexpert}. But at least it is still
being advocated by some experts in most respected journals. For a recent
example with a valuable exposition of the approach and a list of earlier
references see \cite{KiLe97,KiLe99,NaLe00}. A related idea is that of a
conjectured new infrared fixed point in QED$_3$ \cite{AiMa96}.

The present author cannot take a position in the controversy, but he wants to
point out how natural its persistence is: the involved mechanisms include
highly complex interplay of Higgs mechanism, dynamical mass generation (DMG),
confinement and screening. Analytic arguments may miss important points.  This
is at least our experience from the numerical simulation of a lattice model,
the \chup$_3$ model \cite{BaFo98,BaFr98}. This model resembles the 3D
spinon-holon system coupled by the U(1) gauge field, though simplified by
omitting chemical potential used in some cases \cite{KiLe97,NaLe00}.

\section{\chup$_3$ model}

The \chup$_3$ model (see ref.~\cite{BaFo98} for a more detailed description)
is defined on a 3D cubic lattice. The action reads:
\begin{equation}
  \label{chupact}
  S_{\chi U \phi} = S_\chi + S_U + S_\phi,
\end{equation}
with
\begin{eqnarray}
  S_\chi & = & \frac{1}{2} \sum_x \overline{\chi}_x
  \sum_{\mu=1}^3 \eta_{\mu x} (U_{x,\mu} \chi_{x+\mu} \nonumber \\
         &   &  - U^\dagger_{x-\mu,\mu}
  \chi_{x-\mu}) + \hat{m}_0 \sum_x \overline{\chi}_x \chi_x \;,\\
  S_U & = & \beta \sum_{x,\mu<\nu} (1-{\rm Re}\,{U_{x,\mu\nu}}) \;,\\
  S_\phi & = & - \kappa \sum_x \sum_{\mu=1}^3
  (\phi^\dagger_x U_{x,\mu} \phi_{x+\mu} + {\rm h.\,c.}) \;.
\end{eqnarray}

Here $\chi_x$ is the Kogut-Susskind fermion field. Because of doubling our
model describes two four-component fermions ($N_f=2$). We stress that the
charges of the matter fields exclude a direct Yukawa coupling between them.
We are mainly interested in the limit of vanishing bare fermion mass
$\hat{m}_0$ (in lattice units), but allowing nonvanishing $\hat{m}_0$ is
important for better understanding of the model, as well as for technical
reasons. $U_{x,\mu}$ represent the compact U(1) link variables, and
$U_{x,\mu\nu}$ is their plaquette product. The scalar field $\phi$ has, for
simplicity, frozen length $|\phi|=1$. Its hopping parameter $\kappa$ vanishes,
if the square of the bare mass of the scalar field is $+\infty$, and is
infinite, if the bare mass squared is $-\infty$.  Thus large values of
$\kappa$ correspond to the Higgs region, whereas small ones correspond to the
confinement region of the phase diagram.

\begin{figure}
  \begin{center}
    \epsfig{file=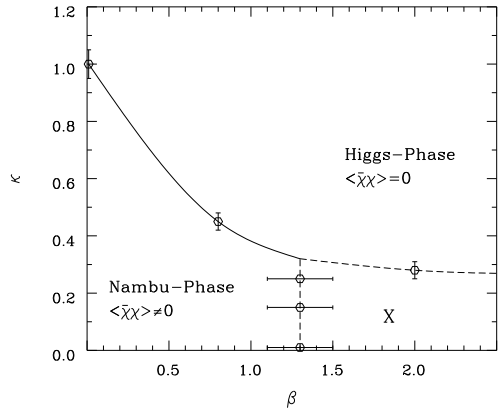,angle=90,width=1.25\hsize}
    \vspace{-10mm}%
    \caption[xxx]{%
      Phase diagram of the \chup$_3$ model at $\hat{m}_0=0$.  (Taken from
      ref.~\cite{BaFo98}.)}
    \label{fig:pd}
  \end{center}
\end{figure}

Our present understanding of the phase diagram in the limit $\hat{m}_0=0$ is shown
in fig.~\ref{fig:pd}. The DMG occurs in the Nambu phase, whereas in the Higgs
phase the fermions remain massless. It is not clear where the boundary between
these phases lies for $\beta>1$. The data is consistent with two possibilities
indicated by dashed lines. Correspondingly, properties of the region denoted
by X are unclear. At least one of the dashed lines must be a phase transition,
one can be a mere crossover.

\section{Limit cases}
For $\beta=0$, the gauge and scalar fields can be integrated out exactly, and
one ends up with a lattice version of a three-dimensional four-fermion model,
the Gross-Neveu or Thirring model (see ref.~\cite{BaFo98} for a discussion of
these alternatives). For our purposes, the important properties of this model
are the second order phase transition at $\hat{m}_0=0, \kappa \simeq 1$, below which
the DMG takes place, and nonperturbative renormalisability in its vicinity,
allowing a continuum limit. At small nonzero values of $\beta$ these
properties persist.

For $\kappa=0$, the scalar field is absent and the model is equivalent to the
compact QED$_3$ with $N_f=2$ fermions. For $\hat{m}_0 \rightarrow \infty$, it
reduces to the pure compact QED$_3$. This is a confining theory, presumably
with some gauge-ball spectrum.

When matter fields are dynamical, various gauge singlets, i.e. unconfined
states are possible, in particular the fermion $F=\phi^\dagger\chi$, which in
the Nambu phase acquires mass through DMG. $F$ would presumably be the
electron in the spinon-holon context.

In the weak gauge coupling limit, $\beta=\infty$, the fermions are free with
mass $\hat{m}_0$, and $S_\phi$ reduces to the XY$_3$ model. It has a phase
transition at $\kappa \simeq 0.27$.

At $\hat{m}_0 = \infty$, the model reduces to the three-dimensional compact
U(1) Higgs model. For its recent investigation and references to earlier
numerical studies see \cite{KaKa98}. Data suggests that the phase transition
of the XY$_3$ model continues as the Higgs phase transition of uncertain order
to finite values of $\beta$. (A different, I think improbable scenario without
the Higgs phase transition has been proposed recently in ref.~\cite{NaLe00}).
At some small $\beta$, there is a critical end point.

The (presumed) knowledge about these limit cases is usually used in analytic
arguments about what happens inside. However, some new phenomena can be
conjectured.
 
\section{A tricritical point?}

The data \cite{BaFo98} suggests that in the interval $0 \le \beta \simeq 0.8$
the phase transition with DMG stays in the same universality class, that of
the four-fermion model. Our first conjecture is that the universality class
nevertheless changes at higher $\beta$, presumably before $\beta \simeq 1.3$,
because a tricritical point may be encountered.

The argument is based on an analogy with a similar model in 4D, the \chup$_4$
model \cite{FrJe95a}. There the line of transitions with DMG meets the lines
of endpoints of the Higgs phase transitions which exist at finite $\hat{m}_0$.
Such a common point of several critical lines (tricritical point) in the
middle of the phase diagram has not been predicted by any analytic approach,
but has been found in a large numerical simulation \cite{FrJe98b}. Tricritical
points are known to be described by universality classes different from those
of any of the entering critical lines.

We expect also in the \chup$_3$ model critical endpoints of the Higgs phase
transitions for finite $\hat{m}_0$. It would be a challenge to find evidence
for them and to check whether they meet the DMG transition line. If so, a new
universality class of DMG would be established in 3D. Its properties might be
qualitatively similar to those found in 4D \cite{FrJe98b}. In particular, the
massive fermion would be accompanied by a massive scalar gauge ball. Could
this be of some interest for the condensed matter physics?

\section{Still another universality class of DMG, \\ 
or a new fixed point in QED$_3$?}

The nature of the region X is of much interest.  Because $\kappa$ is small,
one can neglect the scalar field nearly in the whole region X. Then the
question is what are the properties of compact QED$_3$ at large $\beta$. One
would expect that these are well known. This is not the case, however, because
the perturbation expansion fails to grasp important properties of that theory
even for $\beta \rightarrow \infty$.

On the basis of analytic arguments it is expected that for small $N_F < N_F^c$
the DMG holds in the whole range of $\beta$ including $\beta \rightarrow
\infty$, whereas for $N_F > N_F^c$ it ends at some finite $\beta$. This are
properties similar to QCD$_4$. For noncompact QED$_3$ one expects $ N_F^c
\simeq 3-4$ (see references in \cite{BaFo98}.
\begin{figure}
  \begin{center}
    \epsfig{file=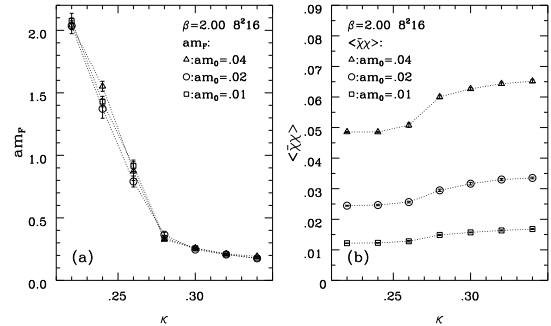,angle=90,width=1.0\hsize}
    \vspace{-10mm}%
    \caption[xxx]{%
      Behaviour of the (a) fermion mass and (b) fermion condensate for
      different small $\hat{m}_0$ as function of $\kappa$ at $\beta=2.00$
      across the horizontal dashed line. (Taken from ref.~\cite{BaFo98}.)}
    \label{fig:fermcbcb200}
  \end{center}
\end{figure}%

In our study of the compact model with $ N_F = 2$, we have found for $\beta
\simeq 1.3$ an indication of a phase transition (vertical dashed line in
fig.~\ref{fig:pd}), which would mean a substantially smaller value of $N_F^c$
than expected in the noncompact case. Naturally, because of the small sizes of
our lattices, we cannot exclude that the condensate rapidly but analytically
decreases around $\beta \simeq 1.3$ to a small but nonvanishing value. It is
very difficult to distinguish numerically such a crossover from a genuine
phase transition.  Therefore the DMG phase transition could also take place on
the horizontal line in fig.~\ref{fig:pd}.

Both alternatives are interesting.  Analogy to noncompact QED$_3$ with
fermions suggests that, for $N_F=2$, DMG might persist until $\beta = \infty$.
Then, provided the transition on the horizontal dashed line in
fig.~\ref{fig:pd} is continuous, a continuum theory with DMG would be obtained
also here. It would contain again the unconfined massive fermion $F$, since
its mass appears to scale (fig.~\ref{fig:fermcbcb200}a).  Thus it would
represent still another universality class with DMG.

The data do really suggest a continuous phase transition on the horizontal
line. However, there is something strange with it: as seen in
fig.~\ref{fig:fermcbcb200}b, the fermion condensate appears to increase with
$\kappa$ across the transition, though, usually, decreasing mass of the scalar
field suppresses this condensate. Such a behaviour requires a clarification.

The analogy to noncompact QED$_3$  (whose properties are far from certain anyhow)
might be misleading, and DMG might end at the vertical dashed line. In this
case, there would be no unconfined fermion of finite mass in the corresponding
continuum theory. However, it would mean a new insight into the properties of
compact QED$_3$, presumably implying the existence of a new fixed point in
this theory, as conjectured e.g. in refs.~\cite{AiMa96}. 

It is known \cite{Po75,GoMa82} that pure compact QED$_3$ has no phase
transition and, as $\beta \rightarrow \infty$, it is confining via a linear
potential.  String tension and a scalar gauge ball mass scale in this limit,
but the scales separate \cite{GoMa82}. Such a rare scale separation
might occur also at the new fixed point of the full QED$_3$.

These alternatives may or may not be of relevance for the condensed matter
physics. But they certainly belong to interesting open questions in
3D gauge theories.

\section{Conclusion}

It is remarkable that after 20 years of numerical studies of lattice gauge
theories so many open questions about the flatland QED with matter fields
remain unanswered. I think the main reason lies in the subtlety of the
problems: to distinguish between crossovers and genuine phase transitions,
between weak first order and second order transitions, to demonstrate scaling
behaviour whose form is not predicted by some reliable analytic means, etc.
Attempts to understand 4D abelian lattice gauge theories have got stuck
because of similar subtleties\cite{Je00}.  In some sense the study of the
QED$_3$ with matter fields is more difficult than the QCD calculations.

The other reason may be a low priority assignment. If so, I hope to have
contributed to its reconsideration.

\vspace{1cm}
{\bf Acknowledgement.}

I am grateful to J. Ho\v{s}ek for drawing my attention to analogies between
our work and some models for high temperature superconductivity, and P.~A.~Lee
for some comments. I thank I.~M. Barbour, E.~Focht, W.~Franzki, { and }\relax
N.~Psycharis for collaboration \cite{BaFo98}, results of which I have used in
this note.

\end{document}